\documentclass[11pt]{article}
\usepackage{amsmath,amssymb,color,graphics,epsfig,cite}
\usepackage{amsfonts}
\newcommand{\hoch}[1]{$\, ^{#1}$}

\makeatletter
\@addtoreset{equation}{section}
\makeatother

\newcommand{\be}{\begin{equation}}
\newcommand{\ee}{\end{equation}}
\newcommand{\bea}{\setlength\arraycolsep{2pt} \begin{eqnarray}}
\newcommand{\eea}{\end{eqnarray}}

\def\ft#1#2{{\textstyle{\frac{\scriptstyle #1}{\scriptstyle #2} } }}
\def\fft#1#2{{\frac{#1}{#2}}}

\def\0{{\sst{(0)}}}
\def\1{{\sst{(1)}}}
\def\2{{\sst{(2)}}}
\def\3{{\sst{(3)}}}
\def\4{{\sst{(4)}}}
\def\5{{\sst{(5)}}}
\def\6{{\sst{(6)}}}
\def\7{{\sst{(7)}}}
\def\8{{\sst{(8)}}}
\def\sst#1{{\scriptscriptstyle #1}}

\def\im{{{\rm i\,}}}

\begin{document}

\begin{flushright}
\hfill{MI-TH-1606}

\end{flushright}

%\vspace{25pt}
\begin{center}
{\Large {\bf Global Scaling Symmetry, Noether Charge}}\\
{\Large {\bf and Universality of Shear Viscosity}}

\vspace{15pt}
{\bf Hai-Shan Liu\hoch{1,2}}

\vspace{10pt}

\hoch{1} {\it Institute for Advanced Physics \& Mathematics,\\
Zhejiang University of Technology, Hangzhou 310023, China}

\vspace{10pt}

\hoch{2} {\it George P. \& Cynthia Woods Mitchell  Institute
for Fundamental Physics and Astronomy,\\
Texas A\&M University, College Station, TX 77843, USA}

\vspace{20pt}

\underline{ABSTRACT}
\end{center}

Recently it was established in Einstein-Maxwell-Dilaton gravity that the KSS viscosity/entropy ratio associated with AdS planar black holes can be viewed as the boundary dual to the generalized Smarr relation of the black holes in the bulk.  In this paper we establish this relation in Einstein gravity with general minimally-coupled matter, and also in theories with an additional non-minimally coupled scalar field.  We consider two examples for explicit demonstrations.

\vfill {\footnotesize Emails: hsliu.zju@gmail.com}

\thispagestyle{empty}

\pagebreak

\tableofcontents
\addtocontents{toc}{\protect\setcounter{tocdepth}{2}}

%%%%%%%%%%%%%%%%%%%%%%%%%%%%%%%%%%%%%%%%

%\newpage
%%%%%%%%%%%%%%%%%%%%%%%%%%%%%%%%%%%%%%%%

\section{Introduction}
The AdS/CFT correspondence\cite{adscft1,adscft2,adscft3,adscft4} has provided many remarkable insights into the connections between gravitational backgrounds in string theory or more general settings and some strongly coupled gauge field theories. Among numerous results established so far,  one well-known example is the universality of the ratio of the shear viscosity to the entropy density for wide classes of gauge theories that have gravity duals, namely \cite{Policastro:2001yc,sonsta,KSS,KSS0}
\be
\fft{\eta}{S}=\fft{1}{4\pi}\,.\label{visent0}
\ee
A number of papers have demonstrated the universality of this bound for a variety of supergravity and gravity theories \cite{Buchel:2003tz,Buchel:2004qq,Benincasa:2006fu,Landsteiner:2007bd,Son:2007vk,Iqbal:2008by} . These proofs are based on two methods, one uses the formula for the viscosity derived from the membrane paradigm\cite{Kovtun:2003wp}, and the other uses Kubo formula\cite{Son:2002sd} to calculate the viscosity which is  proportional to the absorption cross section of a minimal-coupled scalar and the fact\cite{Das:1996we, Emparan:1997iv} that the absorption is equal to the area of the horizon.

Recently, a new method was developed to prove the universality of the ratio of the AdS planar black holes in Einstein-Maxwell-Dilaton theories\cite{liu}.  It was shown that the identity (\ref{visent0}) of the boundary field theory is duel to the bulk generalized Smarr relation.  It turns out that the effective Lagrangian of the AdS planar black holes has a global scaling symmetry.  The generalized Smarr relation is a manifestation of the corresponding Noether charge.  In this paper we extend the proof to more general class of theories that admit planar black holes whose effective Lagrangians have the scaling symmetry. We find that the Noether charge builds a bridge between shear viscosity and the entropy density, which enables us to confirm the universality of the viscosity bound.  On the other hand, evaluating the Noether current on both the horizon and asymptotic infinity in the bulk leads to the generalized Smarr relation \cite{liu}.

The paper is organized as follows. In section 2, we derive the shear viscosity in theory of matter fields minimally coupled to Einstein gravity. Next, we generalize theory to include a non-minimally coupled scalar in section 3. We give two explicit examples in section 4 to confirm our results. The paper ends with conclusions in section 5.

\section{Theories of Einstein Gravity with Minimally Coupled Matter Fields}

   In this section, we give a derivation of the $\eta/s$ ratio
for black holes in the theory of a generic matter field minimally coupled
to Einstein gravity in general dimension $n$.  We require that the effective Lagrangian of the black hole have a global scaling symmetry. The theory is described by the $n$-dimensional action
%%%%%
\be
S_n = \fft1{16\pi G}\, \int d^nx\, \sqrt g \big[ R + M(\phi \,, \nabla \phi \,,g_{\mu\nu}) \big] \,.\label{ndimaction}
\ee
%%%%% We may take the $n$-dimensional
where $M$ denotes the Lagrangian of matter field $\phi$. We shall for now only consider matter field minimally coupled to gravity through $g_{\mu\nu}$,  that $M$ contains no terms that have derivatives on $g_{\mu\nu}$.
%%%%%
The equations of motion are given by
%%%%%
\be
G_{\mu\nu} + M_{\mu\nu} - \fft12 M g_{\mu\nu } = 0 \,, \quad \partial_\mu \Big ( \fft{\sqrt g \delta M}{\delta (\partial_\mu \phi)} \Big) - \sqrt g\Big( \fft{\delta M}{\delta \phi}  \Big) = 0\\,.\label{ESeom}
\ee
%%%%%
where $G_{\mu\nu} = R_{\mu\nu} - \fft12 R g_{\mu\nu}$ is Einstein tensor and $M_{\mu\nu} =\fft{\delta M}{\delta g^{\mu\nu}}$.

\subsection{Isotropic Subspace}

We start by considering the static black-brane with the following general form
%%%%%
\be
ds^2 = dr^2 -a^2 \, dt^2 + b^2 \, dx^i dx^i\,, \quad \phi = \phi(r)  \,, \label{metans}
\ee
%%%%%
where $a$ and $b$ are functions only of $r$.  It should be emphasized that $\phi$ represents a generic matter field, rather than just a simple scalar. The brane subspace $dx^i dx^i$ has an isotropic scaling factor $b^2$. (The anisotropic configuration will be dealt in the next subsection.) Substituting ansatz into (\ref{ESeom}) gives four equations,
%%%%%
\bea
\fft{M_{tt}}{a^2} + \fft12 M - (n-2) \big( \fft{b''}{b} + \fft{n-3}{2} \fft{b'^2}{b^2} \big) = 0\,,\label{ESanseom1}\\
M_{rr} - \fft12 M + (n-2)  \big( \fft{a'b'}{ab} + \fft{n-3}{2} \fft{b'^2}{b^2} \big)  = 0\,,\label{ESanseom2}\\
\fft{M_{xx}}{b^2}- \fft12 M+ \fft{a''}{a} + (n-3) \fft{b''}{b} + \fft{(n-3)(n-4)}{2} \fft{b'^2}{b^2} + (n-3)\fft{a'b'}{ab} = 0\label{ESanseom3}\,,\\
\partial_r \Big ( \fft{\sqrt g \delta M}{\delta \phi'} \Big) - \sqrt g \Big( \fft{\delta M}{\delta \phi}  \Big) = 0\,,\label{ESanseom4}
\eea
%%%%%
We assume that the system admits a black hole solution. In order to see the global scaling symmetry of the system, we substitute the ansatz into the Lagrangian, we have the reduced one-dimensional effective Lagrangian,
%%%%%
\be
 {\cal L} = ab^{n-2} \Big(-\fft{2 a''}{a} - \fft{2 (n-2) b''}{b} -  \fft{(n-2)(n-3)b'^2}{b^2} -\fft{2 (n-2) a'b'}{ab} + M \Big ) \,, \label{lagrangian}
\ee
%%%%%
where the prime denotes a derivative respect to $r$.  The first four terms in the bracket correspond to Einstein gravity and the last term is the matter contribution. We find that the gravity part is invariant under the scaling,
%%%%%
\be
b \rightarrow \lambda b \,, \quad a \rightarrow \lambda^{-(n-2)} a \,. \label{scaling}
\ee
%%%%%
In order for the whole system to have the scaling invariance, we can require the matter field scale correspondingly, namely
%%%%%
\be
\phi \rightarrow \lambda^{c_\phi} \phi \,, \label{scaling}
\ee
%%%%%
where the constant $c_\phi$ is the scaling dimension of matter field $\phi$. The invariance of the full Lagrangian under this scaling implies that
%%%%%
\be
0=\sum_i c_i \psi_i \fft{\delta M}{\delta \psi_i} \equiv c_t g^{tt}\fft{\delta M}{\delta g^{tt}}  + \sum_{i=1}^{n-2} c_i  g^{x_ix_i}\fft{\delta M}{\delta g^{x_ix_i}}  + c_\phi( \phi \fft{\delta M}{\delta \phi} + \phi' \fft{\delta M}{\delta \phi'}) \,,
\ee
%%%%%
with $c_t = 2 (n-2) \,,  c_i = - 2$. Subsituting ansatz (\ref {metans})  into it, we get
%%%%%
\be
\fft{M_{tt} }{a^2} + \fft{M_{xx}}{b^2} + \Delta (\fft{\delta M}{\delta \phi} \phi + \fft{\delta M}{\delta \phi'} \phi') =0  \,, \label{scaling condition}
\ee
%%%%%
where $\Delta = -\fft{c_\phi}{2 (n-2)}$ is a constant related to the scaling dimension of matter field. Since the subspace is uniform, we further require the matter fields have the full rotational symmetry, i.e. $M_{x_i x_j} = M_{xx} \delta_{ij}$, with $x$ one of the spatial direction. This scaling property (\ref{scaling condition}) plays an important role in solving the perturbation equation of motion which will be presented later.

Since the theory has the scaling symmetry, we can derive the corresponding Noether charge by allowing the transformation parameter $\lambda$ to be $r$ dependent, and obtain the Noether charge
%%%%%
\be
 J= a b^{n-2} (\fft{a'}{a} - \fft{b'}{b} - \Delta \fft{\delta M}{\delta \phi'} \phi ) \,. \label{noe}
\ee
%%%%%
   If we consider a black hole solution of the equations (\ref{ESeom}),
with an event horizon located at $r=r_0$, then near the horizon we shall have
the expansions
%%%%%
\be
a(r) = a_1\, [(r-r_0) + a_2\, (r-r_0)^2 +\cdots]\,,\qquad
b(r) = b_0\, + b_1\, (r-r_0)  + \cdots\,.
\label{nearh}
\ee
%%%%%
(We have written $a(r)$ with an overall scale $a_1$, which
is a ``trivial'' parameter, in the sense that it can be absorbed into a
rescaling of the time coordinate $t$.) With standard procedure we can calculate the temperature and entropy  density of the black hole
%%%%%
\be
T = \fft{a_1}{2 \pi} \,, \quad s = \fft{b_0^{n-2}} {4} \,.
\label{thermo}
\ee
%%%%%
Evaluating Noether charge on the horizon(\ref{nearh}) gives
%%%%%
\be
J = 8 \pi Ts \,. \label {hornoe}
\ee
%%%%%
Evaluating $J$ at asymptotic infinity on the other hand yields mass and other conserved quantities.  The conservation of the Noether charge thus gives rise to the generalized Smarr relation \cite{liu}.

We now consider a transverse-traceless metric perturbation in the
$(n-2)$-dimensional space of the planar section, by making the
replacement
%%%%%
\be
dx^i dx^i \longrightarrow dx^i dx^i + 2\Psi\, dx^1 dx^2\,,\label{pert}
\ee
%%%%%
where for the present purpose it suffices to allow $\Psi$ to depend on
$r$ and $t$ only. This field has $O(2)$ symmetry in the $x_1 x_2$ plane.
The linearized equation for $\Psi$ comes from,
%%%%%
\be
G_{x_1x_2}^\1 + M_{x_1x_2}^\1 - \fft12 M^\0 g_{x_1x_2}^\1 = 0 \,.
\ee
%%%%%
Where, $G_{x_1x_2}^\1 \,, M_{x_1x_2}^\1  \,, g_{x_1x_2}^\1$  are linearised terms in $\Psi$ and $M^\0$ is the unperturbed value.
Due to the $O(2)$ symmetry, the perturbation of matter field can be set to zero consistently. Since matter field is minimally coupled, the tensor $M_{\mu\nu}$ defined under (\ref{ESeom}), which is a function of the metric,  should have the following expansion form at the linear order of $\Psi$
%%%%%
\begin{equation}
\left(
  \begin{array}{cc}   %该矩阵一共2列，每一列都居中放置
    M_{x_1x_1} & M_{x_1x_2} \\  %第一行元素
     M_{x_2x_1}  &  M_{x_2x_2}  \\  %第二行元素
  \end{array}
\right)  \, =  \,  M_{xx}\left(                 %左括号
  \begin{array}{cc}   %该矩阵一共2列，每一列都居中放置
    1 & \Psi \\  %第一行元素
    \Psi & 1 \\  %第二行元素
  \end{array}
\right)   + {\cal O} ( \Psi ^2)          \,.
\end{equation}
%%%%%
Then, with background (\ref{metans}) and equation (\ref{ESanseom3}) the linearised equation is given by,
%%%%%
\bea
\Psi'' + \big( \fft{a'}{a} + (n-2) \fft{b'}{b} \big) \Psi'  - \fft{1}{a^2} \fft{d^2\Psi}{dt^2} = 0 \,.
\eea
%%%%%
For a perturbation of the form $\Psi(t,r) = e^{-\im\omega t}\, \psi(r)$,
we therefore have
%%%%%
\be
\psi'' + \big( \fft{a'}{a} + (n-2) \fft{b'}{b} \big) \psi'  + \fft{\omega^2}{a^2} \psi= 0\,.\label{perteq}
\ee
%%%%%
Near horizon(\ref{nearh}), the equation (\ref{perteq}) therefore takes the form
%%%%%
\be
a_1^2 (r-r_0)^2\, \psi'' + a_1^2 (r-r_0)\, \psi'
   + \omega^2\, \psi = 0   \,,
\ee
%%%%%
which can be solved exactly, leading to the ingoing solution
%%%%%
\be
\psi_{\rm in} \propto
\exp\Big[-\fft{\im\omega\, \log(r-r_0)}{a_1}\Big]\,.
\label{psiin}
\ee
%%%%%
(The outgoing solution is obtained by sending $\omega\longrightarrow
-\omega$ in (\ref{psiin}).)

Since in the Kubo formula we only need to know $\psi$ up to the linear order in $\omega$, we can seek the solution for the metric perturbation away from the horizon, in the approximation where $\omega$ is small. We choose an ansatz of the
form
%%%%%
\be
\psi(r) =  \exp\Big[-\fft{\im\omega}{a_1}\,
                  \log\fft{a(r)}{ \tilde a (r)}\Big]\, \Big(1-\im\omega U(r) + {\cal O} (\omega^2)\Big) \label {psires} \,,
\ee
%%%%%
where $\tilde a(r)$ is chosen to make the wave function approches to $1$  at infinity. Keeping terms only up to linear order in $\omega$, we find that
$U(r)$ satisfies the equation
%%%%%
\be
U'' + \big(\fft{a'}{a} + (n-2) \fft{b'}{b} \big) U' + \fft{1}{a_1} \Big(\fft{ a''}{a} +  (n-2) \fft{a'b'}{ab} - \fft{\tilde a''}{\tilde a} + \fft{\tilde a '^2}{\tilde a ^2} - \fft{a'\tilde a'}{\tilde a a} - (n-2) \fft{\tilde a'b'}{\tilde a b} \Big) = 0\,,
\ee
%%%%%
which can be solved by
%%%%%
\be
U'(r) = \fft{1}{a_1} (\fft{\tilde a'}{\tilde a }  - \fft{b'}{b} - \Delta \fft{\delta M}{\delta \phi'} \phi)\,.
\ee
%%%%%
In the following, we shall show that only the expression of $U'(r)$ is needed to calculate the viscosity, the exact expression for $U(r)$ is not necessary.

   We can derive the viscosity by using standard methods described
in the literature.  For our purpose, it is convenient to follow the
procedure given in \cite{sonsta,Brigante:2007nu}, making use of the Kubo formula.  The
first step involves calculating the terms in the action at
quadratic order in the metric perturbation $\Psi(t,r)$.  When doing so, one should
include the Gibbons-Hawking term in the original action. However, one simple way to do so is   to remove the second derivatives on $\Psi$ by performing integrations by parts in the quadratic action.  Then the action at quadratic order has the form
%%%%%
\be
S_n^{(2)} = \fft1{16\pi G}\int d^n x \Big[P_1 \, {\Psi'}^2 +
  P_2\, \Psi\, \Psi' + P_3\, \Psi^2 + P_4\, \dot\Psi^2\Big]\,,
\label{S2}
\ee
%%%%%
with
%%%%%
\be
P_1=-\ft12 r^{n-2}\, ab^{n-2}\,.
\ee
%%%%%
Since the matter field is coupled to gravity through $g_{\mu\nu}$, the derivative terms come from gravity part. The prescription described in \cite{sonsta,Brigante:2007nu}
requires knowing only the $P_1\, \Psi\Psi'$ term, for which we have
%%%%%
\be
\int {\cal L}_n^{(2)}\, dr d^{n-2}x =-
  \ft12\, ab^{n-2}\, \Psi\Psi'\Big|_{r=\infty} \,,\label{surfaction1}
\ee
%%%%%
As only term to linear order of $\omega$ is needed in the calculation of viscosity , with  (\ref{psires}), we expand the surface term to linear order in $\omega$
%%%%%
\be
 -ab^{n-2}\, \Psi\Psi' = \fft{i \omega ab^{n-2}}{a_1} \big( \fft{a'}{a} - \fft{\tilde a '}{\tilde a} + a_1 U'\big) \,.
 \ee
%%%%%
Substituting $U'$, we find the surface term is proportional to Noether charge,
%%%%%
\be
 -ab^{n-2}\, \Psi\Psi'\Big|_{r = \infty} = \fft{i \omega}{a_1} J =  i 4 \omega  s \,.
\ee
%%%%%
Using the prescription in \cite{sonsta,Brigante:2007nu}, we therefore find that
the viscosity is given by
%%%%%
\be
\eta =  \fft{s}{4 \pi}\,. \label{etares}
\ee
%%%%%
It is worthwhile to emphasise at this point that the derivation of  the above universal value  is valid  without needing any specific solution of equation of motion(\ref{ESeom}) or the explicit form of matter field, and hence the result is rather general.

The generalisation to multiple matter fields is straightforward. Here, we shall skip the detailed derivation and just give the key results. The lagrangian  we consider is
%%%%%
\be
{\cal L} = \sqrt g \big( R + M(\Phi_I \,, \nabla \Phi_I \,,g_{\mu\nu}) \big) \,, \label{multi}
\ee
%%%%%
with static metric ansatz and matter fields
%%%%%
\be
ds_{\sst n}^2 = dr^2 - a^2 dt^2 + b^2 (dx^2_1 + ... + dx_i^2) \,, \quad \Phi_I= \Phi_I(r) \,.
\ee
%%%%%
where we use $I$ to denote different matter fields. The lagrangian is invariant under scaling
%%%%%
\be
b \rightarrow \lambda b \,, \quad a \rightarrow \lambda^{-(n-2)} a \,,  \quad \Phi_I \rightarrow \lambda^{c_{\Phi_I}}  \Phi_I \,,
\ee
%%%%%
where $c_{\Phi_I}$ is the scaling dimension of matter field $\Phi_I$. The scaling property of matter field can be expressed as
%%%%%
\be
\fft{M_{tt} }{a^2} + \fft{M_{xx}}{b^2} +\sum_I \Delta_I (\fft{\delta M}{\delta \Phi_I} \Phi_I + \fft{\delta M}{\delta \Phi_I'} \Phi_I') =0 \,, \label{matsca}
\ee
%%%%%
with $\Delta_I = -\fft{c_{\Phi_I}}{2 (n-2)}$.
The corresponding Noether charge is
%%%%%
\be
J= a b^{n-2} (\fft{a'}{a} - \fft{b'}{b} - \sum_I\Delta_I \fft{\delta M}{\delta \Phi_I'} \Phi_I ) = 8 \pi T s \,.
\ee
%%%%%
Doing perturbation (\ref{pert}), the linearised equation of motion for perturbation has the same form as that of previous case(\ref{perteq}). With the same ansatz(\ref{psires}), the equation can be solved to linear order in $\omega$ by
%%%%%
\be
U' =  \fft{1}{a_1} (\fft{\tilde a'}{\tilde a} - \fft{b'}{b} - \sum_I \Delta_I \fft{\delta M}{\delta \Phi_I'}  \Phi_I) \,.
\ee
%%%%%
Then, following the same procedure to calculate viscosity, one gets the universal result
%%%%%
\be
\fft{\eta}{s} = \fft{1}{4 \pi} \,.
\ee
%%%%%

\subsection{Anisotropic Subspace}

In this subsection, we consider Einstein gravity coupled to multiple matter fields(\ref{multi}) with black hole background that has two uniform subspaces
%%%%%
\be
ds_{\sst n}^2 = dr^2 - a^2 dt^2 + b^2 (dx^2_1 + ... + dx_p^2) + c^2 (dy^2_1 + ... + dy_q^2)  \,. \label{metsub2}
\ee
%%%%%
$a\,, b\,, c\,$ and matter fields $\Phi_I \,$ are only functions of $r$ and we shall denote these two subspaces as $x$-space and $y$-space respectively. Since the background can have one more subspace, the lagrangian has two copies of scaling symmetry
%%%%%
\be
b \rightarrow \lambda_x b \,, \quad c\rightarrow \lambda_y c \,,   \quad a \rightarrow \lambda_x^{-p} \lambda_y^{-q}  a \,,  \quad \Phi_I \rightarrow \lambda_x^{c_{\Phi_{xI}}} \lambda_y^{c_{\Phi_{yI}}}  \Phi_I \,.  \label{sca2}
\ee
%%%%%
Where $\lambda_x\,,\lambda_y\,$ are scaling parameters related to $x$-space and $y$-space respectively and $c_{\Phi_{xI}} \,, c_{\Phi_{xI}} \,$  are the corresponding scaling dimension of matter field $\Phi_I$. The matter part has the scaling properties
%%%%%
\bea
\fft{M_{tt} }{a^2} + \fft{M_{xx}}{b^2} +\sum_I \Delta_{xI} (\fft{\delta M}{\delta \Phi_I} \Phi_I + \fft{\delta M}{\delta \Phi_I'} \Phi_I') =0 \,. \cr
\fft{M_{tt} }{a^2} + \fft{M_{yy}}{c^2} + \sum_I\Delta_{yI} (\fft{\delta M}{\delta \Phi_I} \Phi_I + \fft{\delta M}{\delta \Phi_I'} \Phi_I') =0 \,. \label{matter2}
\eea
%%%%%
where $\Delta_{xI} = -\fft{c_{\Phi_{xI}}}{2p} \,, \Delta_{yI} =  -\fft{c_{\Phi_{yI}}}{2q}$  and $M_{xx} \,, M_{yy}$ are the diagonal components of matter tensor in $x$-space and $y$-space respectively. With the similar procedure, one can derive the associated Noether charges
%%%%%
\bea
J_x= a b^p c^q (\fft{a'}{a} - \fft{b'}{b} - \sum_I \Delta_{xI}\Phi_I \fft{\delta M}{\delta \Phi_I'}  )  \,, \cr
J_y= a b^p c^q (\fft{a'}{a} - \fft{c'}{c} - \sum_I \Delta_{yI} \Phi_I \fft{\delta M}{\delta \Phi_I'}  )  \,.
\eea
%%%%%
We consider the black hole background with an event horizon located at $r=r_0$, then near the horizon we have expansions
%%%%%
\be
a(r) \approx a_1[ (r-r_0) + a_2\, (r-r_0)^2]\,, \,
b(r) \approx b_0\, + b_1\, (r-r_0)   \,, \,
c(r) \approx c_0\, + c_1\, (r-r_0)  \,.
\label{nearh2}
\ee
%%%%%
The temperature and entropy density can be calculated through standard method
%%%%%
\be
T = \fft{a_1}{2 \pi} \,, \quad s = \fft{b_0^p c_0^q} {4} \,.
\label{thermo2}
\ee
%%%%%
%%%%%
And evaluating Noether charges on the horizon gives
%%%%%
\be
J_x= J_y = 8\pi Ts  \,.
\ee
%%%%%
Now, we consider a transverse-traceless metric perturbation in the $x$-space , by making the replacement
%%%%%
\be
dx^idx^i \rightarrow dx^idx^i + \Psi dx^1 dx^2  \,.
\ee
%%%%%
Following the steps in the previous section, one can get linearised equation for $\Psi$ from the $x_1x_2$ component of equation of motion. With the same ansatz (\ref{psires}), one can solve the equation by
%%%%%
\be
U' =  \fft{1}{a_1} (\fft{\tilde a'}{\tilde a} - \fft{b'}{b} - \sum_I \Delta_{xI}\Phi_I \fft{\delta M}{\delta \Phi_I'}) \,, \quad \text{or} \quad
U' =  \fft{1}{a_1} (\fft{\tilde a'}{\tilde a} - \fft{c'}{c} - \sum_I \Delta_{yI} \Phi_I \fft{\delta M}{\delta \Phi_I'})
\ee
%%%%%
Combining this and the Noether charge, the viscosity is given by
%%%%%
\be
\eta = \fft{s}{4\pi}  \,. \label{ratio}
\ee
One can also do the perturbation in the $y$-space
%%%%%
\be
dy^idy^i \rightarrow dy^idy^i + \Psi dy^1 dy^2  \,,
\ee
%%%%%
and will finally get the same value as that of (\ref{ratio}).

%%%%%
\section{Including an Non-minimally Coupled Scalar}

In the previous section, we considered theories in which matter fields couple  to gravity minimally.  In this section, we want to go one step further, we add a non-minimally coupled scalar to the previous theory (\ref{multi}), namely
%%%%%
\be
{\cal L} = \sqrt g \big[ \kappa(\phi)R - \fft12 (\partial \phi)^2  - V(\phi) + M(\phi \,,\nabla \Phi_I \,,\Phi_I \,, g_{\mu\nu})  \big]
\ee
%%%%%
for which the equations of motion are
%%%%%
\bea
\kappa (\phi)  (R_{\mu\nu} - \fft12 R g_{\mu\nu}) - \big(\nabla_\mu \nabla_\nu \kappa(\phi) - \Box \kappa(\phi) g_{\mu\nu}\big)
%%%%%
 &-& \fft12 \big( \partial_\mu \phi \partial_\nu \phi -\fft12 (\partial \phi)^2 g_{\mu\nu}\big)   \cr
 %%%%%
& +& M_{\mu\nu} - \fft12 M g_{\mu\nu}  +  \fft12 V  g_{\mu\nu} = 0\,, \cr
\Box \phi - \fft{\partial V(\phi)}{\partial \phi} + \fft{\partial \kappa (\phi)}{\partial \phi} R + \fft{\delta M}{\delta \phi} = 0  \,,
%%%%%
&&  \partial_\mu \Big(\sqrt g \fft{\delta M}{\delta (\partial_\mu \Phi_I')}\Big) - \sqrt g \fft{\delta M}{ \delta \Phi_I} =0 \,.
\eea
%%%%%

We consider general black brane solution and static matter fields
%%%%%
\be
ds^2 = dr^2 - a^2 dt^2 + b^2 dx_i^2 \,, \quad \phi = \phi(r) \,, \quad \Phi_I = \Phi_I(r) \,.
\ee
%%%%%
The lagrangian has a scaling symmetry
%%%%%
\be
b \rightarrow \lambda b \,, \quad a \rightarrow \lambda^{-(n-2)} a \,, \quad \Phi \rightarrow \lambda^{c_{\Phi_I}} \Phi_I \,, \quad \phi \rightarrow \phi \,.
\ee
%%%%%
Note that non-minimally coupled scalar $\phi$ is invariant under this scaling, with scaling dimension equals to zero. The corresponding Noether charge is
%%%%%
\be
J = \kappa ab^{n-2} (\fft{a'}{a} - \fft{b'}{b} - \sum_I \fft{\Delta_I }{\kappa} \fft{\delta M}{\delta \Phi_I} \Phi_I ) \,,
\ee
%%%%%
where $\Delta_I = -\fft{c^{\Phi_I}}{2 (n-2)}$. Notice that $\Phi_I$ appears in the Noether charge, whilst $\phi$ doesn't, since its scaling dimension is zero . So the scaling property of matter part is unchanged in this case, and it takes the same form as that of minimally coupled case (\ref{matsca}).
Near the horizon (\ref{nearh}), we can calculate the temperature and entropy density
%%%%%
\be
T = \fft{a_1}{2 \pi} \,, \quad s = \fft{\kappa_0 b_0^{n-2}} {4} \,,
\ee
%%%%%
where $\kappa_0$ is the value of $\kappa(\phi)$ on the horizon. Since the scalar is non-minimally coupled to gravity, there is an additional factor $\kappa_0$ compared to the minimally coupled case. However, remembering that the Noether charge also has a $\kappa$ factor, it turns out to be that the relationship between Noether charge and entropy density is unchanged. This can be verified by evaluating Noether charge on the horizon
%%%%%
\be
J = 8 \pi Ts \,.
\ee
%%%%%
Then, with similar method, one can do a perturbation, get the linearised equation for the perturbation and find that the equation can solved with the form of (\ref{psires}) by
%%%%%
\be
U' = \fft{1}{a_1} (\fft{\tilde a'}{\tilde a} - \fft{b'}{b} - \sum_I \fft{\Delta_I }{\kappa} \fft{\delta M}{\delta \Phi_I} \Phi_I ) \,.
\ee
%%%%%
Since now the relevant surface term in the quadratic
action of the linearised mode is
%%%%%
\be
-\ft12\,  \kappa(\phi) ab^{n-2}\,
\Psi\Psi'\Big|_{r=\infty} \,,\label{surfaction2}
\ee
%%%%
rather than the previous expression (\ref{surfaction1}), it follows that
the viscosity/entropy ratio will be the same value as the minimally coupled case.   Thus we
see that again $\eta/s= 1/(4\pi)$ through the connection of the
Noether charge.

We can also generalise it to two subspaces(\ref{metsub2}),  the lagrangian is invariant under scaling(\ref{sca2}), the scaling property of matter part is the same as (\ref{matter2}). However, the corresponding Noether charges are changed by a factor of $\kappa$
%%%%
\bea
J_x= \kappa a b^p c^q (\fft{a'}{a} - \fft{b'}{b} - \sum_I \fft{\Delta_{xI}}{\kappa} \fft{\delta M}{\delta \Phi_I'} \Phi_I )  \,, \cr
J_y= \kappa a b^p c^q (\fft{a'}{a} - \fft{c'}{c} - \sum_I \fft{\Delta_{yI} }{\kappa} \fft{\delta M}{\delta \Phi_I'} \Phi_I )  \,,
\eea
%%%%
where $\Delta_{xI} = -\fft{c^{\Phi_{xI}}}{2 (n-2)} \,, \Delta_{yI} = -\fft{c^{\Phi_{yI}}}{2 (n-2)}$. Evaluating on the horizon(\ref{nearh2}) gives
%%%%%
\be
J_x= J_y = 8\pi Ts  \,.
\ee
%%%%%
Then, one can consider a perturbation, and again with form(\ref{psires}), the perturbation equation can be solved by
%%%%
\be
U' = \fft{1}{a_1} (\fft{\tilde a'}{\tilde a}  - \fft{b'}{b} - \sum_I \fft{\Delta_{xI}}{\kappa} \fft{\delta M}{\delta \Phi_I'} \Phi_I )    \quad \text{or} \quad
U' = \fft{1}{a_1} (\fft{\tilde a'}{\tilde a} - \fft{c'}{c} - \sum_I \fft{\Delta_{yI}}{\kappa} \fft{\delta M}{\delta \Phi_I'} \Phi_I ) \,.
\ee
%%%%
Now, the surface term of the quadratic action is
%%%%
\be
-\fft12 \kappa ab^pc^q \psi \psi'\Big|_{r=\infty}  \,.
\ee
%%%%
Making use of Noether charge,  one can again get the viscosity
%%%%
\be
\eta = \fft{s}{4\pi} \,.
\ee
%%%%
The result is not surpring since a non-minimally coupled scalar is invariant under the scaling, acting like a singlet, and doesn't contribute to the Noether charge which connects the viscosity and entropy density.

\section{Explicit Examples}
In the previous sections, we derived the universal result $\eta/s = 1/(4 \pi)$ for general theories of Einstein gravity minimally coupled to matter fields with or without a non-minimally coupled scalar. In this section we study two explicit theories, namely Einstein-Proca and massive vector.  In all cases we find that the scaling property of matter part is satisfied as what we derived in the general theory and the KSS bound is saturated as expect. (The Einstein-Maxwell-Dilaton theory is a specific example of the non-minimally coupled scalar case, which has been studied in \cite{liu}, the result of the viscosity ratio presented there is consistent with ours.)

\subsection{Einstein-Proca gravity}

Einstein-Proca gravity are the simplest theory \cite{taylor} for constructing Lifshitz spacetimes\cite{kachru}.  Lifshitz or AdS black holes in pure Einstein gravity also exists, although no exact solutions were constructed.  Recently the first law of thermodynamics were analysed for these black holes\cite{liulif,Liu:2014tra}.  We now use this as an example to derive the viscosity/entropy ratio for black holes in this theory even though no exact solutions were known. The lagrangian is
%%%%
\be
{\cal L} = \sqrt g (R - 2 \Lambda -\fft14 F^2 -\fft12 m^2 A^2) \,,
\ee
%%%%
where $\Lambda$ is cosmological constant and $F=dA$. Considering the following ansatz,
\be
ds_n^2 = dr^2 - a^2 dt^2 + b^2 dx_i^2 \,, \quad A = \phi dt \,,
\ee
the lagrangian in terms of $a,b,\phi$ is given by,
%%%%
\be
{\cal L} = ab^{n-2} \Big(-\fft{2 a''}{a} - \fft{2 (n-2) b''}{b} -  \fft{(n-2)(n-3)b'^2}{b^2} -\fft{2 (n-2) a'b'}{ab} + \fft{m^2 \phi^2+\phi'^2}{2 a^2} - 2 \Lambda\Big) \,.
\ee
%%%%
The lagrangian is invariant under scaling
%%%%
\be
b\rightarrow\lambda b \,, \quad a \rightarrow \lambda^{-(n-2)} a \,, \quad \phi \rightarrow \lambda^{-(n-2)} \phi \,,
\ee
%%%%
with $c_\phi = - (n-2)$. And the corresponding Noether charge is
%%%%
\be
J = ab^{n-2} \big( \fft{a'}{a} - \fft{b'}{b} - \fft{\phi \phi'}{2 a^2} \big) \,
\ee
%%%%
which equals to $8\pi Ts$ when evaluating on the horizon. One can verify that the vector field has the scaling property (\ref{scaling condition}) with $M=-\fft14 F^2-\fft12 A^2$ and $\Delta = \fft12$.

Then, one can do the perturbation, get the linearized equation for the perturbation and solve the equation with form(\ref{psires}) by
%%%%
\be
U' = \fft{1}{a_1} (\fft{\tilde a'}{\tilde a}  - \fft{b'}{b} - \fft{\phi \phi'}{2 a^2}) \,.
\ee
%%%%
Combining this and the Noether charge, one can calculate the viscosity
%%%%
\be
\eta = \fft{s}{4 \pi} \,.
\ee
%%%%

\subsection{Massive p-Form}
For examples with anisotropic brane configurations, we consider Einstein gravity coupled to a massive $(p-1)$-form potential:
%%%%
\be
 {\cal L} = \sqrt g \Big( R - 2\Lambda - \fft{1}{2 p!} F^2_{\sst{ (p)}} - \fft{m^2}{2 (p-1)!} A^2_{\sst{(p-1)}} \Big) \,,
\ee
%%%%
where $\Lambda$ is cosmological constant and $F_\sst {(p)} = dA_{(p-1)}$. We consider the black brane metric ansatz with two subspaces and static form
%%%%
\bea
ds_{\sst n}^2 &=& dr^2 - a^2 dt^2 + b^2 (dx^2_1 + ... + dx_{p-2}^2) + c^2 (dy_1^2 + ....+ dy_{n-p}^2) \,, \cr
A &=& \phi dt\wedge dx_1\wedge ... \wedge dx_{p-2} \,.
\eea
%%%%
As observed before, there are two copies of scaling symmetry
%%%%
\be
b\rightarrow \lambda_x b \,, \quad c\rightarrow \lambda_y c \,, \quad a\rightarrow \lambda_x^{-(p-2)} \lambda_y^{-(n-p)} a \,, \quad \phi \rightarrow \lambda_y^{-(n-p)}\phi \,.
\ee
%%%%
The corresponding Noether charges are
%%%%
\bea
J_x&=& a b^{p-2} c^{n-p} (\fft{a'}{a} - \fft{b'}{b})  \,, \cr
J_y &=& a b^{p-2} c^{n-p} (\fft{a'}{a} - \fft{c'}{c} - \fft{\phi\phi'}{2 a^2 b^{2 (p-2)}})  \,.
\eea
%%%%
Note that the form field has scaling dimension zero under scaling related to $x$-space, so the form field doesn't appear in the charge related to  $x$-space. One can verify that the two scaling properties (\ref{matter2})  for $p$-form are satisfied.

Following the similar procedure, one can do the perturbation, get the linearized equation for the perturbation and solve the equation with form(\ref{psires}) by
%%%%
\be
U' = \fft{1}{a_1} (\fft{\tilde a'}{\tilde a}  - \fft{b'}{b} )  \quad \text{or} \quad
U' = \fft{1}{a_1} (\fft{\tilde a'}{\tilde a}   - \fft{c'}{c} - \fft{\phi\phi'}{2 a^2 b^{2 (p-2)}}) \,.
\ee
%%%%
Combining this and the Noether charge, one can get the viscosity
%%%%
\be
\eta = \fft{s}{4 \pi} \,.
\ee
One can see that, when set ${b=c\,, p=2}$, the scaling degree of $x$-space vanishes, and the system goes back to Einstein-Proca case.

%%%%

\section{Conclusion}

In this paper we considered the planar black holes that have a global  scaling symmetry. Such a scaling symmetry does not exist for spherically-symmetric black holes.  We focused on the theories with some generic minimally-coupled matter fields, and derived the scaling properties for these matter fields. We then show that system has a conserved Noether charge which equals to $8\pi T s$ when evaluated on the horizon.  We used the Kubo formula to compute the viscosity and found that the result was related to the Noether charge, and hence confirmed the universality of the KSS viscosity/entropy ratio.  The Noether charges in these black holes are related to the generalized Smarr relation, which is derived by evaluating the Noether charge both on the horizon and on the asymptotic infinity.  Our results demonstrate a duality relation between the universality viscosity/entropy ratio of the boundary theory and the generalized Smarr relation on the bulk.  It should be emphasized that our proof breaks down for black holes that do not exhibit such a globle scaling symmetry and indeed the viscosity/ratio bound can be
violated in such cases \cite{hartnoll}.

Our results are general, for planar black holes with isotropic brane subspace or general anisotropic subspaces.  We further extend the conclusion to include a non-minimally coupled scalar field as well.  The scaling symmetry and the corresponding Noether charge play an important role in our derivation, whilst the existence of an analytical black hole solution is of less important.  This technique may be useful in other linear response system in the AdS/CFT correspondence.

\section*{Acknowledgements}

We are grateful to Hong Lu and Chris Pope for reading the manuscript. The work is supported in part by DOE Grant No. DE-FG02-13ER42020, NSFC grants No. 11305140, 11375153 and 11475148,  and CSC scholarship No. 201408330017.


\begin{thebibliography}{99}

\bibitem{adscft1} J.M. Maldacena,
{\it The large N limit of superconformal field theories and supergravity},
Adv.\ Theor.\ Math.\ Phys.\  {\bf 2}, 231 (1998),
hep-th/9711200.
%%CITATION = HEP-TH 9711200;%%bm{mal}

\bibitem{adscft2} S.S. Gubser, I.R. Klebanov and A.M. Polyakov,
{\it Gauge theory correlators from non-critical string theory},
Phys.\ Lett.\  {\bf B428}, 105 (1998),
hep-th/9802109.
%%CITATION = HEP-TH 9802109;%%

\bibitem{adscft3} E. Witten,
{\it Anti-de Sitter space and holography},
Adv. Theor. Math. Phys. {\bf 2}, 253 (1998), hep-th/9802150.
%%CITATION = HEP-TH 9802150;%%

\bibitem{adscft4}
  O. Aharony, S.S. Gubser, J.M. Maldacena, H. Ooguri and Y. Oz,
{\it Large $N$ field theories, string theory and gravity,}
  Phys.\ Rept.\  {\bf 323}, 183 (2000)
  [hep-th/9905111].
  %%CITATION = HEP-TH/9905111;%%
  %3507 citations counted in INSPIRE as of 29 Aug 2015

%\cite{Policastro:2001yc}
\bibitem{Policastro:2001yc}
  G. Policastro, D.T. Son and A.O. Starinets,
{\it The shear viscosity of strongly coupled ${\cal N}=4$
supersymmetric Yang-Mills plasma,}
  Phys.\ Rev.\ Lett.\  {\bf 87}, 081601 (2001),
 hep-th/0104066.
  %%CITATION = HEP-TH/0104066;%%
  %968 citations counted in INSPIRE as of 06 juil. 2015

\bibitem{sonsta} D.T. Son and A.O. Starinets,
{\it Minkowski space correlators in AdS/CFT correspondence:
Recipe and applications},
JHEP {\bf 0209}, 042 (2002), hep-th/0205051.
%%CITATION = HEP-TH/0205051;%%

\bibitem{KSS} P. Kovtun, D.T. Son and A.O. Starinets,
{\it Holography and hydrodynamics: Diffusion on stretched horizons},
JHEP {\bf 0310}, 064 (2003), hep-th/0309213.
%%CITATION = HEP-TH/0309213;%%

\bibitem{KSS0} P. Kovtun, D.T. Son and A.O. Starinets,
{\it Viscosity in strongly interacting quantum field theories from black
hole physics},  Phys.\ Rev.\ Lett.\  {\bf 94}, 111601 (2005),
hep-th/0405231.
%%CITATION = HEP-TH/0405231;%%


%\cite{Buchel:2003tz}
\bibitem{Buchel:2003tz}
  A.~Buchel and J.~T.~Liu,
{\it Universality of the shear viscosity in supergravity,}
  Phys.\ Rev.\ Lett.\  {\bf 93}, 090602 (2004),
  hep-th/0311175.
  %%CITATION = HEP-TH/0311175;%%
  %354 citations counted in INSPIRE as of 06 juil. 2015

%\cite{Buchel:2004qq}
\bibitem{Buchel:2004qq}
  A.~Buchel,
{\it On universality of stress-energy tensor correlation functions in
supergravity,}
  Phys.\ Lett.\ B {\bf 609}, 392 (2005),
  hep-th/0408095.
  %%CITATION = HEP-TH/0408095;%%
  %138 citations counted in INSPIRE as of 06 juil. 2015

%\cite{Benincasa:2006fu}
\bibitem{Benincasa:2006fu}
  P.~Benincasa, A.~Buchel and R.~Naryshkin,
{\it The shear viscosity of gauge theory plasma with chemical potentials,}
  Phys.\ Lett.\ B {\bf 645}, 309 (2007),
  hep-th/0610145.
  %%CITATION = HEP-TH/0610145;%%
  %72 citations counted in INSPIRE as of 06 Jul 2015

%\cite{Landsteiner:2007bd}
\bibitem{Landsteiner:2007bd}
  K.~Landsteiner and J.~Mas,
{\it The shear viscosity of the non-commutative plasma,}
  JHEP {\bf 0707}, 088 (2007),
  arXiv:0706.0411 [hep-th].
  %%CITATION = ARXIV:0706.0411;%%
  %41 citations counted in INSPIRE as of 06 juil. 2015

%\cite{Son:2007vk}
\bibitem{Son:2007vk}
  D.~T.~Son and A.~O.~Starinets,
  {\it Viscosity, Black Holes, and Quantum Field Theory,}
  Ann.\ Rev.\ Nucl.\ Part.\ Sci.\  {\bf 57}, 95 (2007)
  [arXiv:0704.0240 [hep-th]].
  %%CITATION = doi:10.1146/annurev.nucl.57.090506.123120;%%
  %412 citations counted in INSPIRE as of 03 Jan 2016

%\cite{Iqbal:2008by}
\bibitem{Iqbal:2008by}
  N. Iqbal and H. Liu,
{\it Universality of the hydrodynamic limit in AdS/CFT and
the membrane paradigm,}
  Phys.\ Rev.\ D {\bf 79}, 025023 (2009),
  arXiv:0809.3808 [hep-th].
  %%CITATION = ARXIV:0809.3808;%%
  %297 citations counted in INSPIRE as of 06 juil. 2015

  %\cite{Kovtun:2003wp}
\bibitem{Kovtun:2003wp}
  P.~Kovtun, D.~T.~Son and A.~O.~Starinets,
 {\it Holography and hydrodynamics: Diffusion on stretched horizons,}
  JHEP {\bf 0310}, 064 (2003)
  [hep-th/0309213].
  %%CITATION = doi:10.1088/1126-6708/2003/10/064;%%
  %444 citations counted in INSPIRE as of 03 Jan 2016

  %\cite{Son:2002sd}
\bibitem{Son:2002sd}
  D.~T.~Son and A.~O.~Starinets,
  {\it Minkowski space correlators in AdS / CFT correspondence: Recipe and applications,}
  JHEP {\bf 0209}, 042 (2002)
  [hep-th/0205051].
  %%CITATION = doi:10.1088/1126-6708/2002/09/042;%%
  %714 citations counted in INSPIRE as of 03 Jan 2016

%\cite{Das:1996we}
\bibitem{Das:1996we}
  S.~R.~Das, G.~W.~Gibbons and S.~D.~Mathur,
  {\it Universality of low-energy absorption cross-sections for black holes,}
  Phys.\ Rev.\ Lett.\  {\bf 78}, 417 (1997)
  [hep-th/9609052].
  %%CITATION = doi:10.1103/PhysRevLett.78.417;%%
  %211 citations counted in INSPIRE as of 03 Jan 2016

%\cite{Emparan:1997iv}
\bibitem{Emparan:1997iv}
  R.~Emparan,
  {\it Absorption of scalars by extended objects,}
  Nucl.\ Phys.\ B {\bf 516}, 297 (1998)
  [hep-th/9706204].
  %%CITATION = doi:10.1016/S0550-3213(98)00076-5;%%
  %40 citations counted in INSPIRE as of 03 Jan 2016

  %\cite{Liu:2015tqa}
\bibitem{liu}
  H.~S.~Liu, H.~Lu and C.~N.~Pope,
   {\it  Generalized Smarr formula and the viscosity bound for Einstein-Maxwell-dilaton black holes,}
  Phys.\ Rev.\ D {\bf 92}, no. 6, 064014 (2015)
  [arXiv:1507.02294 [hep-th]].
  %%CITATION = doi:10.1103/PhysRevD.92.064014;%%
  %8 citations counted in INSPIRE as of 21 janv. 2016

  %\cite{Brigante:2007nu}
\bibitem{Brigante:2007nu}
  M.~Brigante, H.~Liu, R.~C.~Myers, S.~Shenker and S.~Yaida,
  {\it Viscosity Bound Violation in Higher Derivative Gravity,}
  Phys.\ Rev.\ D {\bf 77}, 126006 (2008)
  [arXiv:0712.0805 [hep-th]].
  %%CITATION = doi:10.1103/PhysRevD.77.126006;%%
  %297 citations counted in INSPIRE as of 03 Jan 2016


%\cite{Taylor:2008tg}
\bibitem{taylor}
  M.~Taylor,
  {\it Non-relativistic holography},
  arXiv:0812.0530 [hep-th].
  %%CITATION = ARXIV:0812.0530;%%
  %315 citations counted in INSPIRE as of 21 janv. 2016

  %\cite{Kachru:2008yh}
\bibitem{kachru}
  S.~Kachru, X.~Liu and M.~Mulligan,
  {\it Gravity duals of Lifshitz-like fixed points,}
  Phys.\ Rev.\ D {\bf 78}, 106005 (2008)
  [arXiv:0808.1725 [hep-th]].
  %%CITATION = doi:10.1103/PhysRevD.78.106005;%%
  %584 citations counted in INSPIRE as of 21 janv. 2016


  %\cite{Liu:2014dva}
\bibitem{liulif}
  H.~S.~Liu and H.~L\"u,
  {\it Thermodynamics of Lifshitz Black Holes,}
  JHEP {\bf 1412}, 071 (2014)
  [arXiv:1410.6181 [hep-th]].
  %%CITATION = doi:10.1007/JHEP12(2014)071;%%
  %17 citations counted in INSPIRE as of 21 Jan 2016

%\cite{Liu:2014tra}
\bibitem{Liu:2014tra}
  H.S.~Liu, H.~L\"u and C.N.~Pope,
{\it Thermodynamics of Einstein-Proca AdS black holes,}
  JHEP {\bf 1406}, 109 (2014)
  [arXiv:1402.5153 [hep-th]].
  %%CITATION = doi:10.1007/JHEP06(2014)109;%%
  %13 citations counted in INSPIRE as of 26 Jan 2016


%\cite{Hartnoll:2016tri}
\bibitem{hartnoll} 
  S.~A.~Hartnoll, D.~M.~Ramirez and J.~E.~Santos,
 {\it Entropy production, viscosity bounds and bumpy black holes,}
  arXiv:1601.02757 [hep-th].
  %%CITATION = ARXIV:1601.02757;%%
  %2 citations counted in INSPIRE as of 27 Jan 2016


\end{thebibliography}
\end{document}